\documentclass[preprint,12pt]{elsarticle}
\usepackage{amssymb}
\usepackage{color,soul} % for highlighting
\usepackage{caption}
\usepackage{subcaption}
\usepackage{lineno}
\journal{Nuclear Instruments and Methods A}

\begin{document}

\begin{frontmatter}



\title{Incident Beamline Design for a Modern Cold Triple Axis Spectrometer at the High Flux Isotope Reactor\tnoteref{fnCPY}}
\tnotetext[fnCPY]{Notice: This manuscript has been authored by UT-Battelle, LLC, under contract DE-AC05-00OR22725 with the US Department of Energy (DOE). The US government retains and the publisher, by accepting the article for publication, acknowledges that the US government retains a nonexclusive, paid-up, irrevocable, worldwide license to publish or reproduce the published form of this manuscript, or allow others to do so, for US government purposes. DOE will provide public access to these results of federally sponsored research in accordance with the DOE Public Access Plan (https://www.energy.gov/doe-public-access-plan)}.

\author[ornl]{G. E. Granroth}
\author[gatech]{M. Daum}
\author[ornl]{A. A. Aczel}
\author[ornl]{T. J. Williams\fnref{fn1}}
\fntext[fn1]{Current affiliation: ISIS Neutron and Muon Source, STFC Rutherford Appleton Laboratory, Didcot OX11 0QX, United Kingdom}
\author[ornl]{B. Winn}
\author[ornl]{J. A. Fernandez-Baca}
\author[gatech]{M. Mourigal}
\author[ornl]{M. D. Lumsden}

\affiliation[ornl]{organization={Neutron Scattering Division},%Department and Organization
            addressline={Oak Ridge National Laboratory}, 
            city={Oak Ridge},
            postcode={37831}, 
            state={Tennessee},
            country={USA}}
\affiliation[gatech]{organization={School of Physics},%Department and Organization
            addressline={Georgia Institute of Technology}, 
            city={Atlanta},
            postcode={30332}, 
            state={Georgia},
            country={USA}}

\begin{abstract}
A modern cold triple axis spectrometer is being planned for the High Flux Isotope Reactor (HFIR) at Oak Ridge National Laboratory. 
Here, we describe the design of an incident beamline that will put a flux of $\sim 10^8\mathrm{\frac{n}{cm^2 s}}$ on a sample with an area of 2 cm x 2 cm. 
It takes current physical constraints at HFIR into account and it can accommodate both single and multiplexed analyzer-detector secondary spectrometers and large superconducting magnets. 
The proposed incident beamline includes a multi-channel guide with horizontal focusing, a neutron velocity selector, components to facilitate an incident beam polarization option, and a double-focusing pyrolytic graphite monochromator. 
This work describes the process of optimizing the guide system and monochromator and summarizes the expected performance of the incident beamline for non-polarized operation.

\end{abstract}



\begin{keyword}
Neutron instrumentation,
Neutron beam modeling,
Neutron simulations,
Triple axis spectrometer



\end{keyword}

\end{frontmatter}


\section{Introduction}
The High Flux Isotope Reactor (HFIR) at Oak Ridge National Laboratory is planning to update the guide system that transports neutrons to the cold guide hall instruments. 
A modern cold triple axis spectrometer (TAS) will be installed at one of the end-guide positions.
To ensure optimal performance, a design of the incident beamline that takes the physical constraints of this location into account has been performed.

The main goal of this design study is to ensure that the cold triple axis spectrometer will have a high flux comparable to world-class instruments at other neutron scattering facilities around the world. The key science requirement is to achieve a maximum neutron fluence on a 2 cm $\times$ 2 cm spot for an incident energy ($E_i$) range from 2.4 meV to 20 meV.
Though specifying an upper optimization range we are not specifying a sharp cutoff.  The instrument will be able to operate at higher energies,  but the instrument is not optimized for such operation. 

This criteria will also ensure significant performance overlap with the HB-1 and HB-3 thermal triple axis spectrometers at the facility. The high flux of this instrument will help to facilitate a polarization analysis capability. 
Several polarization modes are planned, including single-axis, XYZ \cite{93_scharpf}, spherical neutron polarimetry \cite{93_brown}, and ultra-fine resolution measurements using Wollaston prisms \cite{doi:10.1063/1.4875984, Li:in5021}. 

With these instrument parameters, facility constraints, and through surveying recent TAS designs \cite{Rodriguez_2008,SKOULATOS2011100,KOMAREK201163,Piovano_2014,doi:10.1080/10448632.2015.1057050,SCHMALZL201689,UTSCHICK201688,CHENG201617}, one can envision a guide system that provides a virtual source to feed a double-focusing pyrolytic graphite (PG) monochromator which reflects a monochromatic beam on the sample. 
A velocity selector will be added to the incident beam upstream of the monochromator to remove higher-order wavelength contamination and a V-channel polarizer on a translation table is required to ensure both unpolarized and polarized incident beam options. 
In this work, the instrument design of the unpolarized incident beam option for this cold triple axis spectrometer is described. 
The geometry and reflectivity coating optimizations of the guide system are discussed first, the monochromator design is discussed next, and finally the performance of the neutron velocity selector is presented. Generally flux on sample was the figure of merit used for characterizing the performance.  
In a couple of cases other parameters are used in addition to better distinguish the best contribution. 

The optimization and performance characterization rely heavily on the Monte Carlo Ray tracing simulations using the McStas code \cite{Willendrup:2021vp}. 
Monte Carlo Ray tracing sends probability packets, originating from a source, through a series of components to a detector. 
Each component modifies the probability of a packet according to the physics of that component. 
This provides a way to model the complex behavior of multiple components by only specifying the physics of individual components.
The probability packets are scaled according to the output of the source, so as more packets are sent through the system, a more statistically significant estimate of the fluence at a given position along the beam is calculated.
Since these packets roughly approximate individual neutrons, they are described in this more intuitive way throughout this work.   

 Simulations with different statistics were performed and a level of $1\times 10^9$ neutrons was found to provide sufficient detail required for the optimization.
Multiple types of monitors were placed at various locations along the beamline simulation to characterize the performance of an individual configuration.
These are a set of monitors that show the beam profile, the horizontal and vertical acceptance, and the wavelength dependence of the neutron packets.
These were placed at the end of the guide, at the virtual source, just before the monochromator, and at the sample position. Analysis of the results from these monitors was performed with the mcstasclasspy tools \cite{mc_class}.

For all simulations a white neutron beam with wavelengths as long as 10~\AA~was produced by the source component. This ensured that a single simulation could be used to explore the wavelength-dependence of the incident beamline design upstream of the monochromator. For obtaining simulation results at the sample position, separate simulations were performed for each $E_i$. While these simulations could in principle be run faster by narrowing the incident wavelength bandwidth, a broad bandwidth was still used so higher-order wavelength rejection by the velocity selector could be tested.
 
\section{Guide Design}
\label{GuideDesign}
The neutron guide design is largely controlled by the constraints placed on it by the source and the pre-existing buildings at HFIR.
The components upstream of the shutter are common to all the instruments in the cold guide hall. The optimization of those components and the shutter guide is described elsewhere \cite{MFrost2021}.
These components were used to feed the guide that is optimized from here out.
Besides the aforementioned description upstream of the shutter, the guide system must follow a narrowly defined path from the shutter to the cold guide hall using existing openings in the building shielding\cite{CROW2011S71}.
This path was designed to block streaming of high energy neutrons and $\gamma$-rays from the reactor core for personnel protection and background reduction.
It is also important to note that the beam path points slightly up. The monochromator position, which is determined by the available space in the cold guide hall, controls where the guide ends. To identify the optimal monochromator position, the space required for the longest instrument configuration (i.e. the Wollaston prism set-up) was considered. For these Larmor-type experiments to achieve similar energy resolution to TRISP \cite{doi:10.1080/10448630701328372} at the Heinz Maier-Leibnitz Zentrum, the Wollaston prisms in both the incident and scattered beam should be about 55~cm long \cite{FankangDiscussion}. Accounting for the Wollaston prism specification and ensuring that there is $\sim 1$ m clearance between the instrument and the cold guide hall walls places the monochromator at 28.0~m from the source. The guide system will transport the neutrons to a virtual source located 1.6~m upstream of the monochromator

In addition to the building constraints, other assumptions have been made in order to optimize the instrument and they are now detailed.
First, symmetric Rowland geometry \cite{willis2017experimental} will be used for the horizontal focusing of the monochromator.
This means the distance from the virtual source to the monochromator is the same as the distance from the monochromator to the sample.

Next the focal length must be chosen, In principle one wants a short distance to maximize the flux on sample.  
However ensuring the largest sample environment can be used at the lowest incident energy is a more stringent constraint.  
Specifically 1.6 m was chosen based on the lowest incident energy of 2.4 meV, a sample table of 1~m diameter, and the distance from the guide center to the outside of the incident beamline shielding of 0.71 m. 
This last distance was based on the amount of shielding needed for existing instruments in the HFIR cold guide hall. 
The 2.4 meV energy gives a $2\theta$ value for the monochromator of $\sim 120^\circ$.  With these values a focal length of 1.6 m will allow the sample table to clear the shielding for an $\mathrm{E_i}=2.4$ meV with 175 mm to spare. 

This places the sample 1.6~m downstream from the monochromator and the virtual source 1.6~m upstream of the monochromator. Next, higher-order wavelength contamination will be removed from the incident neutron beam with a neutron velocity selector (NVS) placed as far upstream in the guide hall as possible, which fixes the distance between the NVS and the monochromator at 6~m. 
This choice constrains the guide's cross section to be no larger than the 47~mm wide by 130~mm tall upstream window of the standard NVS used at the HFIR\cite{bertzold1989}. 
Furthermore, the divergence through the NVS should be minimized to reduce loss in this device. The placement of the NVS then fixes the dimensions of an horizontal elliptically shaped focusing section of guide between the NVS and the virtual source and a vertical elliptically shaped expanding section from the shutter guide to the NVS. 
With these constraints in place, the shape of the guide system in the horizontal and vertical directions is optimized first, followed by the optimization of the coatings.

The optimization strategy is as such.  The guide coatings are fixed to be m=7 in the horizontal and m=5 in the vertical following the reflectivity curves of Swiss Neutronics\cite{Swiss_neutronics} and then the shape is varied. These coatings are high enough that all neutrons that will be used in coating optimization steps will be propagated by the shape.  So this allows separation of the shape and the coating optimization steps. The above constraints are tight, allowing for little variation in the overall shape. Therefore the channel optimization was performed first. The only parameters to vary are the length of the straight section and the curvature of the guide as they are constrained to fit in the overall length and thread the streaming blocks in the shielding. 
The ellipses are controlled by the NVS and the goal to have the longest focusing length possible so no section of guide has too extreme of a reflection angle\cite{SCHANZER200463,JANOSCHEK2010119}.
   For the coating optimization a first step is performed by examining the maximum reflection angle at each step along the guide.  This allows us to greatly reduce the coating from the initial m= 5 and m=7 values and gets us close to a configuration that ends up with the best performance for the lowest m value coatings. A second step was then performed by examining sections of the guide with high m or with a large area of supermirror to see if the m values could be further reduced. After these two steps, the guide had equivalent performance with significantly lower m values.   

\begin{figure}
    \centering
    \includegraphics[width=0.55\textwidth]{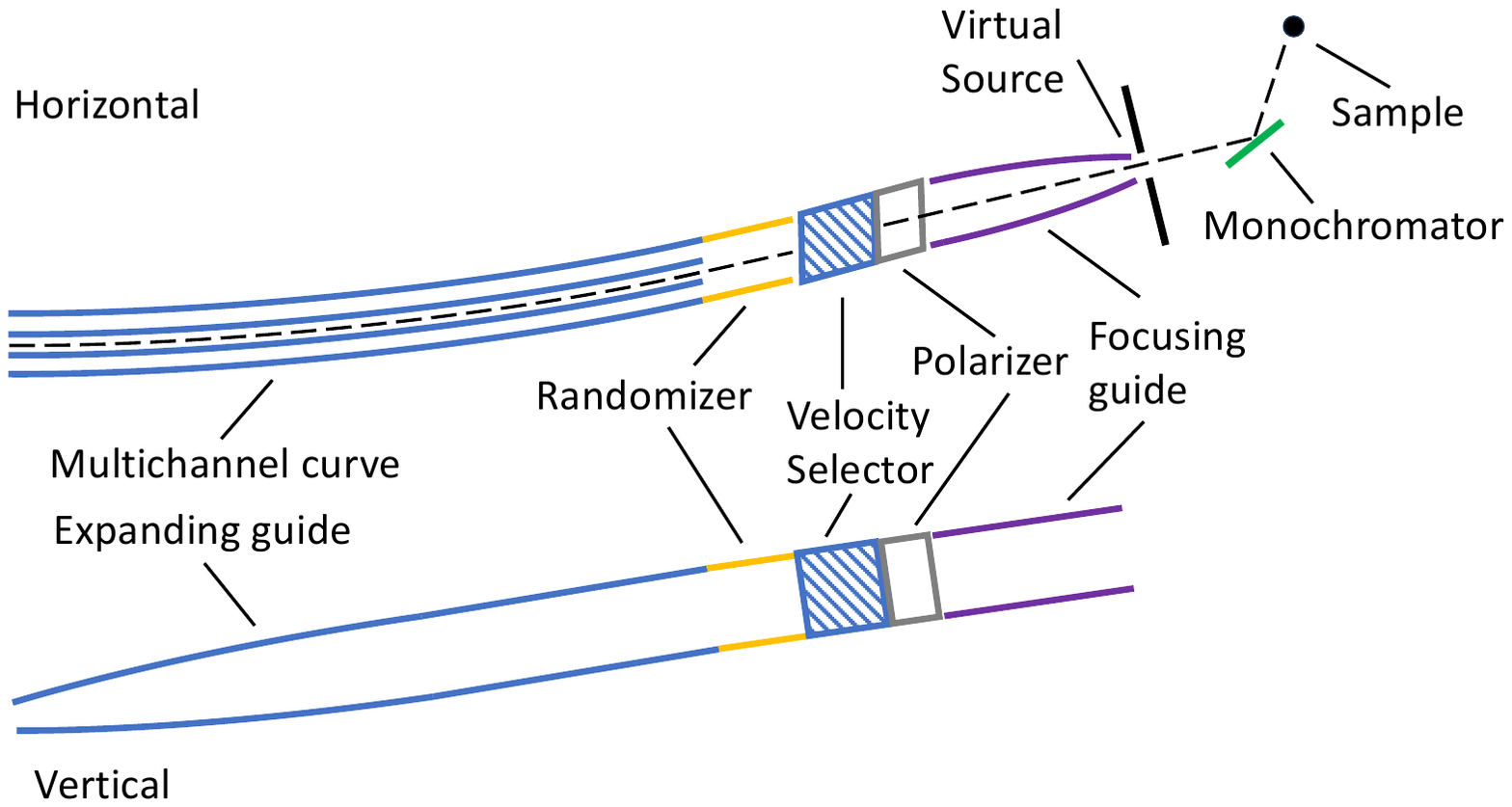}
    \caption{Schematic view of the horizontal and vertical projections of the incident beamline.  
    Each component is labeled. The dashed black line shows the beam path in the center of the horizontal.}
    \label{fig:schematic}
\end{figure}

To help understand the instrument configuration, a schematic showing the various components of the incident beamline is shown in Fig. \ref{fig:schematic}.

 \subsection{Guide shape design}
For the horizontal direction of the guide, there were two optimizations performed. Assuming a guide with a radius of curvature of 275 m, the number of channels in the curve was optimized. Then with that optimization, the length of a randomizing straight section, the radius of curvature of the guide, and the shape of the elliptical focus toward the virtual source were all refined together. Initial attempts to use a compact beam bender or a logarithmic spiral \cite{Hilton1998-oz}, rather than a continuous radius, were tried. However due to the tight constraints of the guide path, no better solution than the curved option was found. 
 
\begin{figure}
    \centering
    \includegraphics[width=1.0\textwidth]{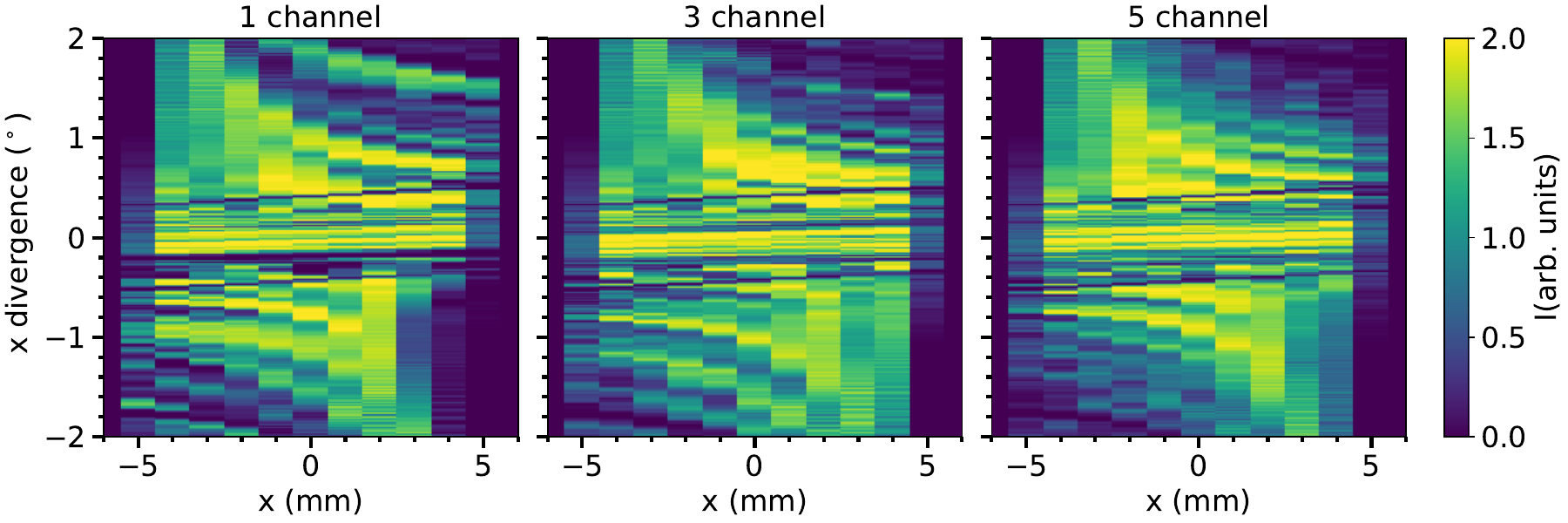}
    \caption{The horizontal phase space at the virtual source for the 1, 3 and 5 channel cases. For the single channel case, there is a gap at $-0.2^\circ$ that is persistent across the whole virtual source width. This gap fills in by increasing the number of channels.}
    \label{fig:VS_acc}
\end{figure}

\begin{figure}   
    \centering
    \includegraphics[width=0.45\textwidth]{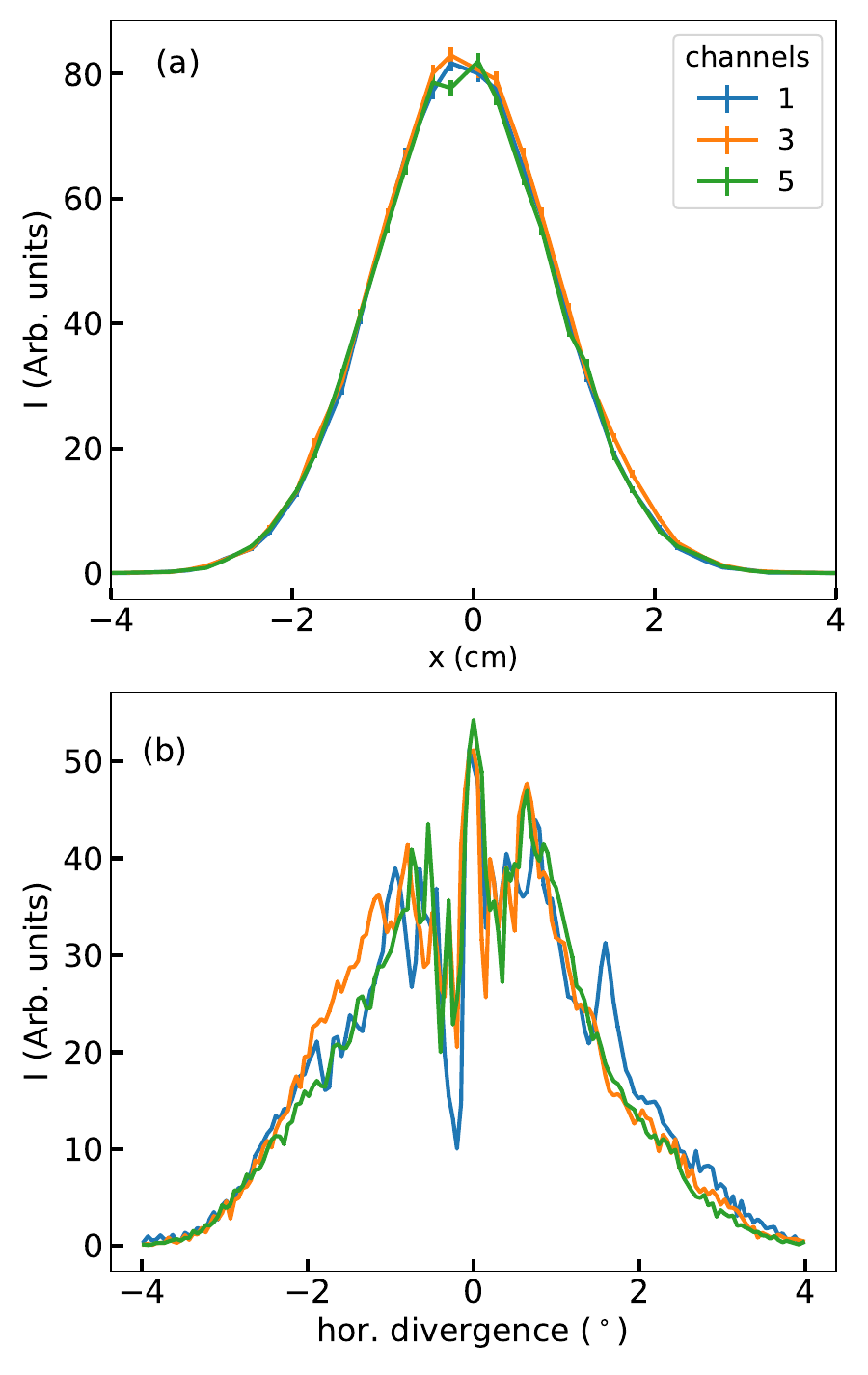}
    \caption{(a) Neutron position distribution in the horizontal direction for a 1~cm tall beam on the sample with 1, 3, or 5 channels in the guide. 
    (b) The horizontal divergence distribution at the virtual source for neutrons in a bandwidth $3.5\pm 2.5$ \AA\, range cut on the sample for 1, 3, or 5 channels in the guide. By introducing multiple channels the sharp dip near $-0.2^\circ$ is mitigated.}
    \label{fig:ChanComp}
\end{figure}

Using the initial radius of curvature, McStas simulations were run using $1\times 10^9$ neutrons with 1, 3, and 5 channels in the curve.
For the 3 and 5 channel cases, the septa between the channels were assumed to be 0.5 mm thick. 
The neutron phase space distribution and the flux as a function of position were monitored at the virtual source and at the sample position. 
For this optimization, it is particularly instructive to watch the divergence distribution at the virtual source and the position distribution on the sample.
The position distribution tells us that the spot is centered on the samples and provides a flux on sample.  The phase space at the virtual source allows us the check the smoothness of the divergence distribution as it will effect the resolution.  Though the phase space will largely be randomized by the monochromator, gaps in the divergence may combine with the focusing monochromator in non-intuitive ways potentially creating a structured resolution function. Having as smooth of a distribution as possible ensures that the resolution volume will be uniform. Figure \ref{fig:VS_acc} shows the horizontal phase space at the virtual source for the 1, 3, and 5 channel cases. 
In all cases there is significant structure, but of particular concern is the gap near $-0.2^\circ$ for the 1 channel case that runs across the horizontal of the virtual source.
We investigated this further and found it is there at all wavelengths. Such a feature was demonstrated in a symmetric fashion for an elliptical snout alone\cite{JANOSCHEK2010119}.
The curve fills this dip in on the positive divergence side and expands it on the negative side.
From there we found it was straightforward to quantify the effect in this feature by taking the divergence summed over a wavelength band of $3.5\pm 2.5$\AA. 
Figure \ref{fig:ChanComp}(b) shows the results for the 1, 3 and 5 channels showing there is significant improvement in the dip near $-0.2 ^\circ$ from 1 to 3 channels, but minimal improvement by moving to 5 channels.
Figure~\ref{fig:ChanComp}(a) shows the neutron position distribution in the horizontal direction for a 1 cm tall beam on the sample and shows that there is little change in flux on sample and no change in width.
 Only minor gains are observed when moving to 5 channels and a small dip in peak flux is observed. Therefore, this aspect of the design is considered optimized with 3 channels in the bender.
 
 With the number of channels confirmed, the next step is to refine the radius of curvature to optimize the randomizing straight section and the elliptical focusing section.
 Since the source and monochromator position are fixed, lengthening the straight section makes the distance to use elliptical focusing toward the virtual source smaller.
 Adding a straight section after a curved guide is well known to make a more uniform beam\cite{RL-79-084,doi:10.13182/NSE92-A23870}. This occurs by bouncing neutrons multiple times in the guide. More bounces ensures there is more mixing, or randomizing, of the phase space distribution. As the primary purpose of this straight section is to randomize the distribution we call it a ``randomizer'' throughout the manuscript.
Ideally  the center of the ellipse starts at the end of the randomizing section and the focal point is at the virtual source. 
However the beam must fit through the opening in the NVS.
This constraint forces the elliptical focusing to be closer to the ellipse focal point than the center for the configuration with the longer randomizing section.  
Though this means there is a discontinuity between the randomizing section and the elliptical focusing, this is better than the effects of having a shorter randomizing section.

\begin{figure}
 \includegraphics[width=0.5\textwidth]{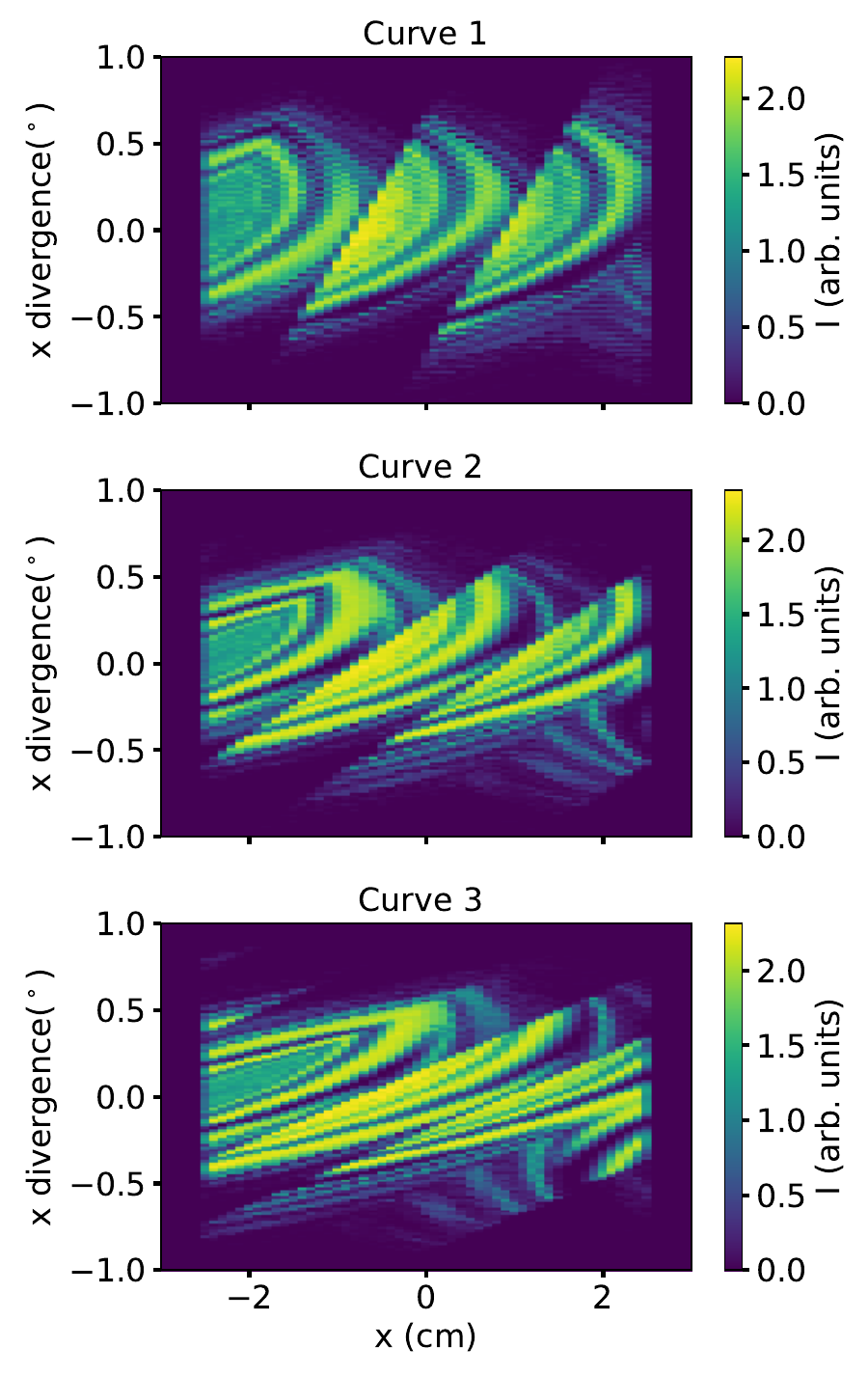}
 \caption{The acceptance diagrams at the end of the guide for the three configurations described in the text and summarized in Table \ref{tab:CurveConfig}.}
 \label{fig:CurveAccDiag}
 \end{figure}

\begin{figure}
 \includegraphics[width=0.45\textwidth]{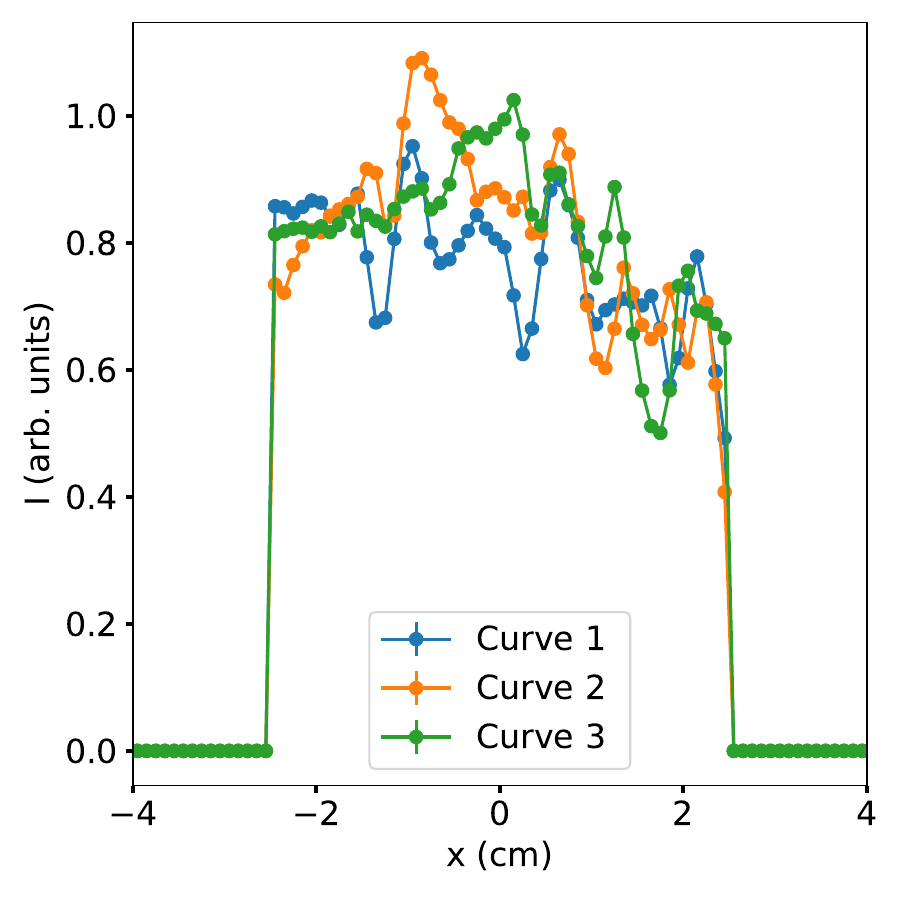}
 \caption{The beam profile just upstream of the NVS  for the three configurations described in the text and summarized in Table \ref{tab:CurveConfig}.}
 \label{fig:Curvexint}
 \end{figure}

 Three different configurations were compared and their parameters are summarized in Table~\ref{tab:CurveConfig}.
 \begin{table}
 \begin{tabular}{|c|c|c|}
 \hline
 Configuration & radius of & randomizing\\
  &curvature (m) & length (m)\\
 \hline 
 Curve 1 &374.9 &0.7 \\
 \hline
 Curve 2 &352.3 & 1.65 \\
 \hline
 Curve 3 &346.4& 2.65 \\
 \hline
 \end{tabular}
 \caption{Comparison of the radius of curvatures and randomizing guide lengths of the three different guide configurations.}
 \label{tab:CurveConfig}
 \end{table}
 Figure \ref{fig:CurveAccDiag} shows acceptance diagrams at the end of the elliptical focusing section of the guide for the three conditions.  
 Three instances of the characteristic half moon shape produced by curved guides \cite{doi:10.1080/10238169308200242} are observed, one for each of the three channels. Notice as the randomizing section is extended, the three half moons are stretched. This stretching allows for better mixing of neutrons coming out of the curve. The improved mixing can be seen by looking at the horizontal profile across the front window of the NVS as shown for the three configurations in Fig.~\ref{fig:Curvexint}. Specifically, configuration 3 shows less pronounced dips and peaks near the center of the beam. The focusing monochromator will further randomize the beam making it more uniform at the sample position. 

After the straight section is the NVS and in the horizontal direction we use a 1/2 ellipse with its center (widest width) just after the NVS and its focal point at the virtual source. The width at the NVS and the virtual source position then fix the shape of the ellipse.  As this optimization is for a 2cm by 2cm sample, we ended the guide when its size was 2 cm wide. Further optimization will be performed to determine if a fixed opening with slits after it is suitable for samples of different sizes or if interchangeable guide options are required for the last section of guide. 
 
For the vertical direction in the guide, the direct view of the source is $0.4^\circ$ above the horizontal. To ensure the vertical shape would not adversely affect the upper energy cutoff, a straight path was chosen in this direction. The beam center will be brought back to the horizontal plane using the monochromator. The initial configuration for the optimization in this direction maximized the guide height at the opening. However as the source is so far back from the guide entrance, the guide is not effectively filled. Furthermore, the guide could be expanded to 130 mm at the velocity selector to minimize the divergence of the neutrons at the monochromator. 
Therefore, an elliptical expanding design is used in the vertical direction.  Here the position of the NVS sets the center of the ellipse and the focal point is set so the ellipse is the size of the source at the source position. Specifically the focus is 1 m upstream of the source and the guide starts after the shutter with a 99.8 mm 
 height expanding to 130 mm in the vertical direction at the NVS. As compared to a straight guide of uniform dimension, this design enhances the flux at shorter wavelengths. Specifically it expands the phase space in distance and shrinks it in divergence allowing the guide coating to be of a lower m value. If we assume a fixed guide coating, it makes the guide less sensitive to angular alignment errors.

\subsection{Guide coating optimization}
\begin{figure}
 \includegraphics[width=0.55\textwidth]{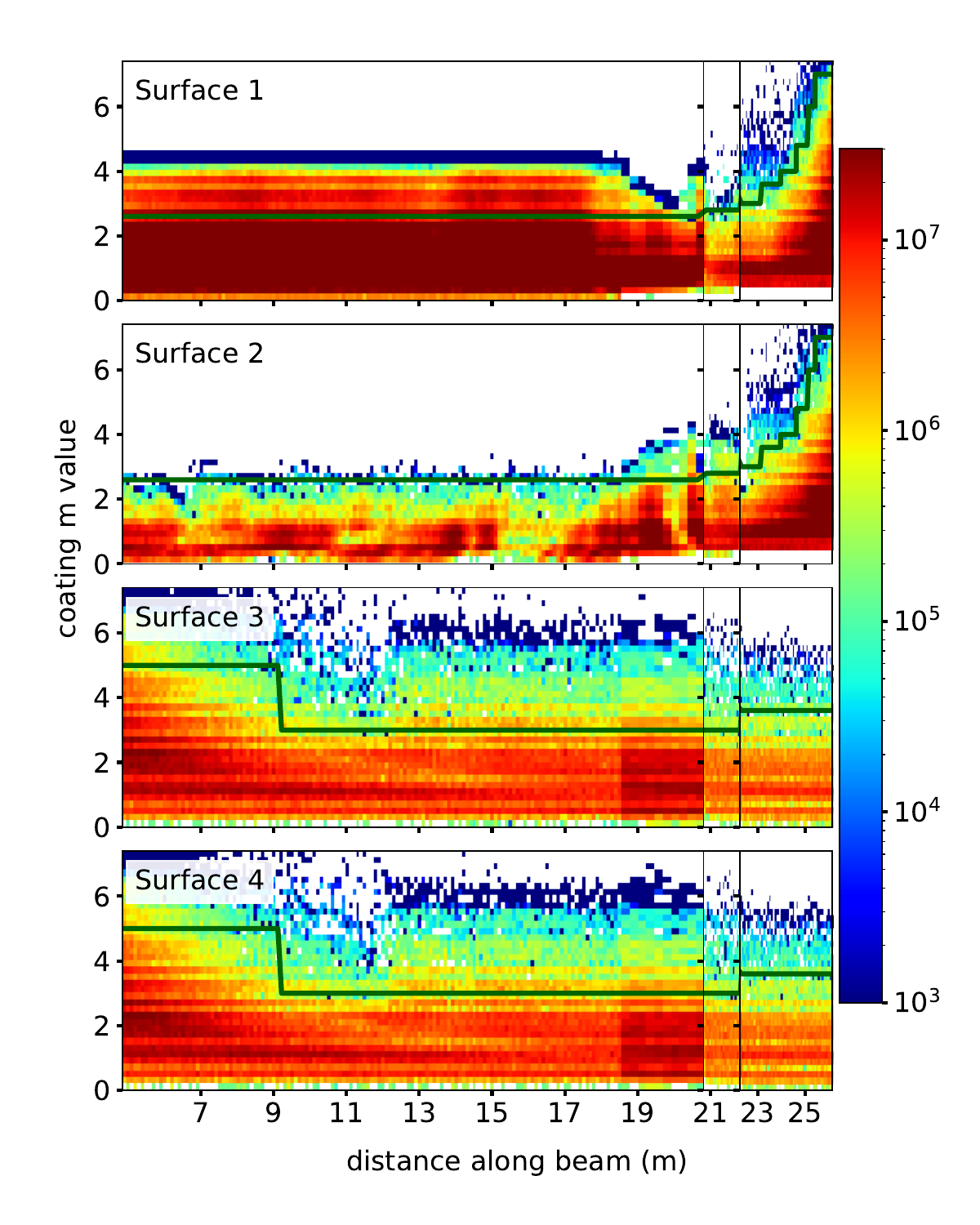}
 \caption{The guide coating histograms provided by the tally components. The dark green curve is the most efficient profile that provides little loss in flux on sample. For clarity of how one guide section affects another, all guide sections are shown together. The first line just below 21 m indicates where guide described by individual components transitions to guide described by off files. 
 The next black line indicates where the velocity selector is located.}
 \label{fig:mhistdist}
 \end{figure}

In the guide shape design simulations described above, the coating values $m$ were set to 7 and 5 in the horizontal and vertical directions respectively. Therefore, all neutrons that could be used were propagated down the system. The challenge of guide optimization is to minimize the guide coating without reducing performance. 
One could try changing the guide coating on each component until a performance hit is observed. However, a more efficient approach uses the the ORNL-developed tally components for McStas, to provide an upper limit on the the guide coating \cite{Huegle_20XX}.
The tally components provide an extension to the McStas probability packet so that each time a neutron reflects off a surface, the angle of reflection, the wavelength of the neutron, and the probability of reflection are recorded. 
This event list is then recorded at the end of the guide. 
Note that this simulation is performed with a 0-10 \AA ~bandwidth provided by the reactor source feeding the guide system.  These events were processed to provide a histogram of intensity versus $m$ for each section of guide. The histograms are summarized as a false color plot of $m$ as a function of distance from the origin of the instrument in Fig.~\ref{fig:mhistdist}.
Surfaces 1 and 2 are for the outer and inner curve of the guide, respectively.  Surfaces 3 and 4 are for the top and bottom surfaces. 
The cutoff in $m$, above which no neutrons are observed, provides an initial set of $m$ values for further optimization.
As the color scale in Fig.~\ref{fig:mhistdist} covers several orders of magnitude, the cutoff line can be moved into the colored region (at least into the blue-green part) with little loss in flux. 

\begin{figure}
 \includegraphics[width=0.45\textwidth]{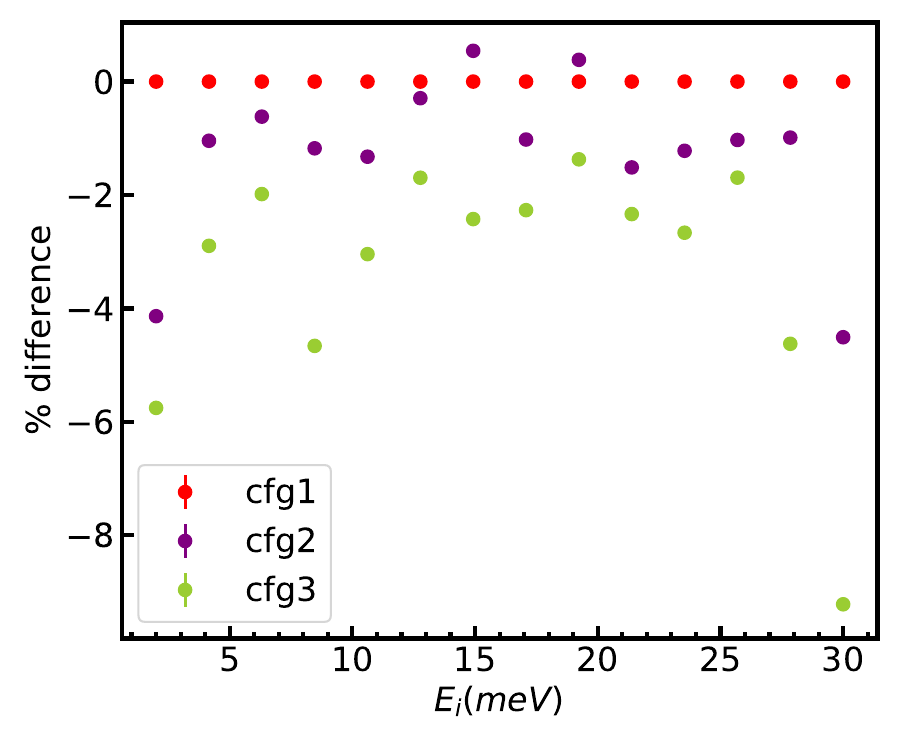}
 \caption{The percent difference for the coating configurations as compared to the initial configuration provided by the tally component method.}
 \label{fig:mcoat}
 \end{figure}

A subsequent set of simulations were used to understand this cutoff. A set of simulations at discrete incident energies was performed and the flux on sample analyzed. Each point was simulated for $1 \times 10^9$ neutrons. Several configurations were tried and three are described here. 
The ultimate point of this exercise is to minimize cost. Though a detailed estimate of the cost is outside of the scope of this work the cost is driven by the amount of area coated and the number of layers in the coating.  This last driver is a function of the m value \cite{10.1117/12.448076}. 
Therefore regions of the guide with large quantities of a coating (e.g. the bender) or where there was high-index guide were targeted for optimization.
Further optimization could be performed, but it would not significantly change the cost drivers. The configurations highlighted here are: the starting configuration (Cfg1), the optimized one (Cfg2), and one that sees an observable performance hit (Cfg3). The surface area for each $m$ value in each configuration is summarized in Table~\ref{tab:surfm} and the percent loss in flux on sample for configurations 2 and 3, as compared to configuration 1, are shown in Fig.~\ref{fig:mcoat}.

\begin{table}
\begin{tabular}{|c|c|c|c|}
\hline
m&Cfg1 area &Cfg2 area & Cfg3 area  \\
   & $(\mathrm{cm^2})$ & $(\mathrm{cm^2})$ & $(\mathrm{cm^2})$ \\
   \hline
2.6& & &104170.237 \\
\hline
2.8& &2547.483&2547.483 \\
\hline
3.0& 1784.270&120480.176& 16309.939 \\
\hline
3.6&2609.616& 4810.955& 4810.955 \\
\hline
3.8& 123651.007& 1138.2& 1138.2 \\
\hline
4.0&1320.120& 2166.691 &2166.691  \\
\hline
4.8 & & 1344.015 &1344.015 \\
\hline
5.0&6409.507&4273.464&4273.464 \\
\hline
6.0& &810.657&810.657 \\
\hline
7.0&3883.005&2085.884&2085.884 \\
\hline
\end{tabular}
\caption {The surface area of each $m$ guide coating required for the three configurations.}
\label{tab:surfm}
\end{table}

The final guide coating profile is provided by Cfg2, the dark green curve in Fig.~\ref{fig:mhistdist}. 
 
\section{Monochromator Design}
To maximize the flux on sample, a double focusing monochromator is envisioned.  This monochromator will operate with variable focusing in both the horizontal and vertical directions. 
This not only allows optimal energy-dependent focusing, but also allows the monochromator to be used in a flat condition when the user wants the finest possible Q-resolution.
Traditional focusing is used in the vertical direction \cite{shirane2002neutron} and a symmetric Rowland design \cite{willis2017experimental}, with a 1.6 m focal length, is used for the horizontal direction. The simulation takes $E_i$ as an input and adjusts the horizontal and vertical focusing of the monochromator accordingly. As a starting point, a monochromator consisting of a 29 $\times$ 29 PG crystal array with crystal dimensions of 12~mm $\times$ 12~mm, horizontal and vertical crystal mosaics of 27', and perfect reflectivity was assumed and used in simulations for the guide optimization. This subsequent monochromator optimization examined its overall size and the size, mosaic, and thickness of individual crystals. 

\begin{figure*}
    \centering
    \includegraphics[width=0.8\textwidth]{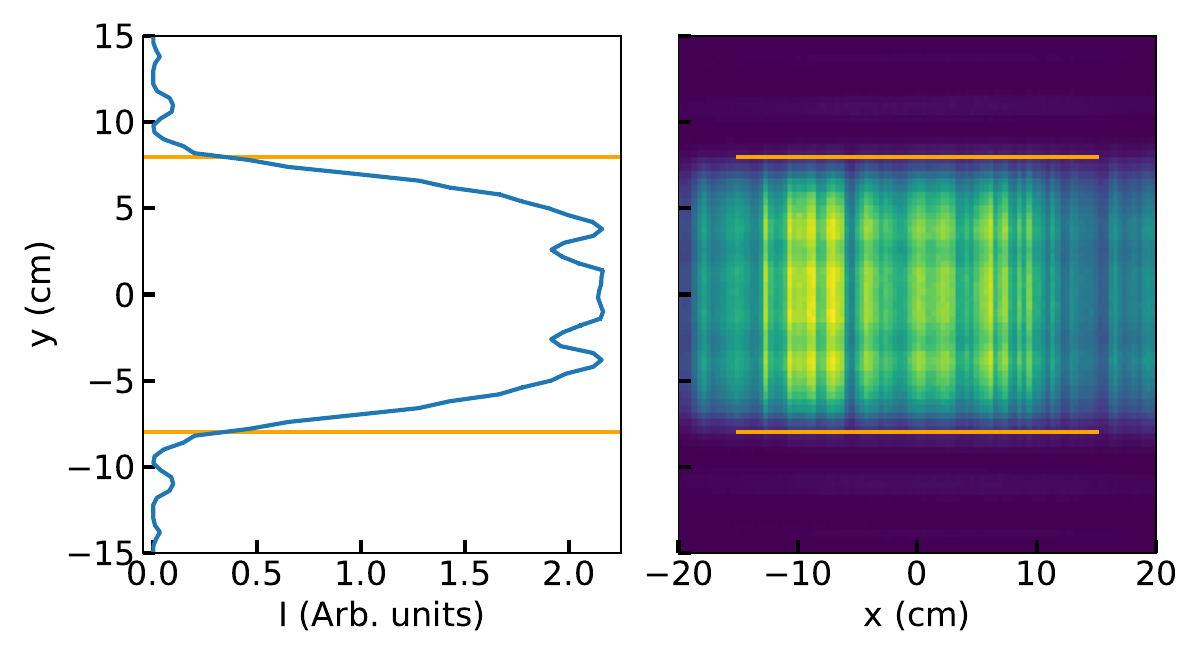}
    \caption{ The figure on the right is a view of the neutrons on the monochromator. The chosen incident energy is 19.23 meV which means the 2$\theta$ angle of the monochromator is $17.9^\circ$. The orange lines indicate the height required for less than a 2\% loss. A cut along the y-direction with a 5 cm width in x is shown on the left. }
    \label{fig:mono_yht}
\end{figure*}

First to study the monochromator size, a two-dimensional image monitor component was placed just in front of the monochromator and rotated with it, so the beam size on the monochromator could be monitored. Simulations were run for $1\times 10^9 $ neutrons over the full energy range. By looking at any specific incident energy value, it is clear that the monochromator can be reduced to 14 crystals high with little loss in flux. Figure~\ref{fig:mono_yht} provides an image view just before the monochromator for an $E_i$ of 19.23 meV, which means the angle of the monochromator is $17.9^\circ$.

\begin{figure}
\includegraphics[width=0.45\textwidth]{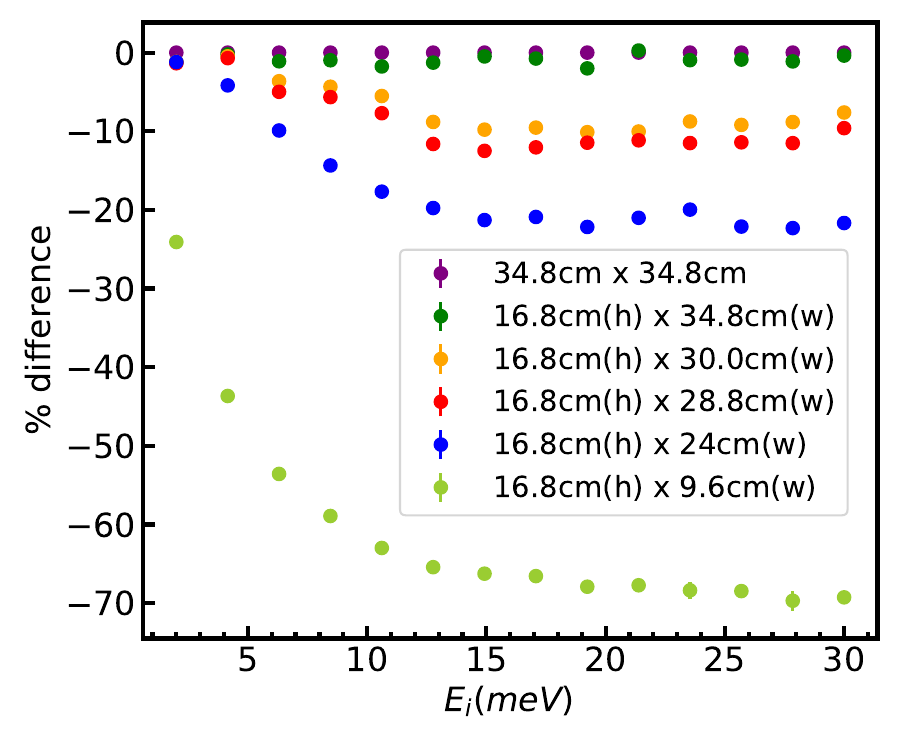}
\caption{The ratio of the flux at the sample position for a smaller monochromator as compared to the initial 34.8 cm $\times$ 34.8 cm design. The 16.8cm (h) x 30.0 (w) cm size (orange points) is the chosen configuration.}
\label{fig:mono_size}
\end{figure}

The orange lines indicate a vertical regime where $>98\%$ of all neutrons interact with the monochromator. Clearly it is straightforward to choose this cutoff. However in the horizontal direction, the choice is not so clear. At the highest energies, the monochromator intersects the beam at a shallow angle and thus a very large crystal array would be required to fully capture the beam. To understand this effect a series of simulations, with varying numbers of monochromator crystals in the horizontal direction, were performed. The results of this study are shown in Fig.~\ref{fig:mono_size}. The purple and green curves show that no loss is observed by reducing the vertical size. The light green curve shows the flux loss incurred if the existing CTAX monochromator is used and provides a performance value which we definitely want to exceed. Notice that decreasing the monochromator size to 24 cm wide means a flux loss of nearly 20\% at an $E_i = 20$ meV. The chosen size is 30 cm wide (orange curve) as this keeps the flux loss to no more than 10\%.

Next the size of the monochromator crystals that make up the assembly was studied.
For perfect focusing, the desire is to have more smaller crystals. 
However to practically make a mechanical focusing mechanism, there are two aspects that favor larger crystals. First, there must be gaps between the crystals.  For the simulations, gaps of 0.5 mm were used. Clearly the crystal size should be sufficiently large that these gaps are negligible.  Second, a focusing mechanism with fewer crystals has fewer moving parts  and is therefore simpler to build. Therefore, the optimization is to increase the size of the crystals without degrading performance. 
These issues are best addressed by ensuring that each row in the monochromator array has an odd number of crystals to avoid a gap at the monochromator center and selecting crystals with dimensions no larger than the desired beam size at the sample position. These last two constraints limit the amount of optimization that could be performed. 
Therefore the beam spot at the sample position for a crystal size of 15 mm $\times$ 15 mm was compared to the initial value of 12 mm $\times$ 12 mm. Neither the beam height nor the FWHM were adversely affected by such a change. 

\begin{figure*}
    \centering
    \includegraphics[width=0.8\textwidth]{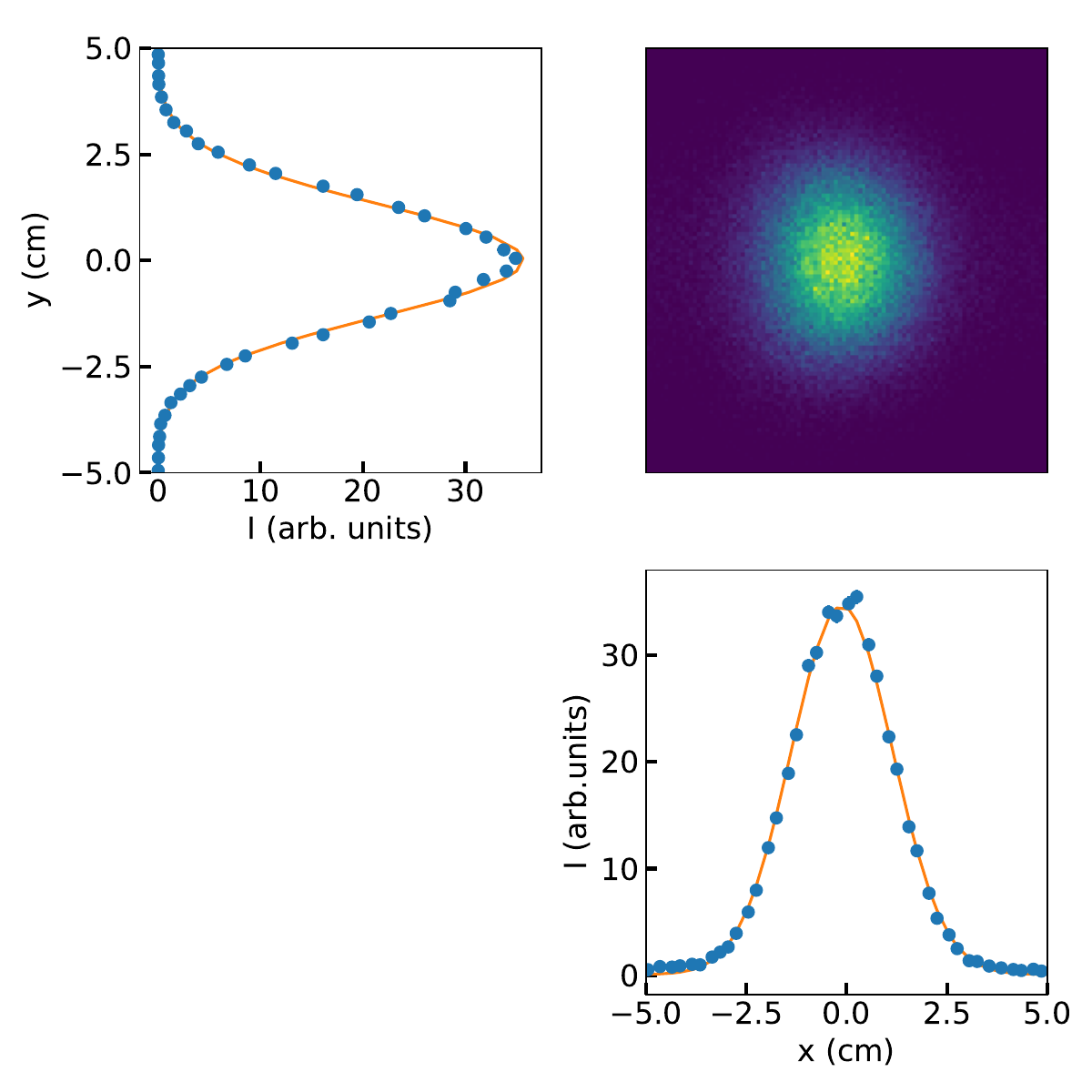}
    \caption{A view of the spot monitor at the sample position for ${E_i} =$~6.31 meV and a crystal mosaic of 27' is provided in the upper right. Below that is a cut averaged over a 2 cm band around the center in the y direction plotted against x.  The plot to the upper right is a cut averaged over a 2 cm band around the center in the x direction plotted against y. The orange curve is the best fit to a Gaussian to characterize the spot size.}
    \label{fig:spot_xmp}
\end{figure*}

Next the effect of the crystal mosaic was studied. Three values were compared, including 27', 36', and 48'. For each of these mosaic choices, a beam was propagated to a beam image monitor at the sample position for several incident energy settings. Figure~\ref{fig:spot_xmp} shows an example of the result at the sample position. 

\begin{figure}
    \centering
    \includegraphics[width=0.5\textwidth]{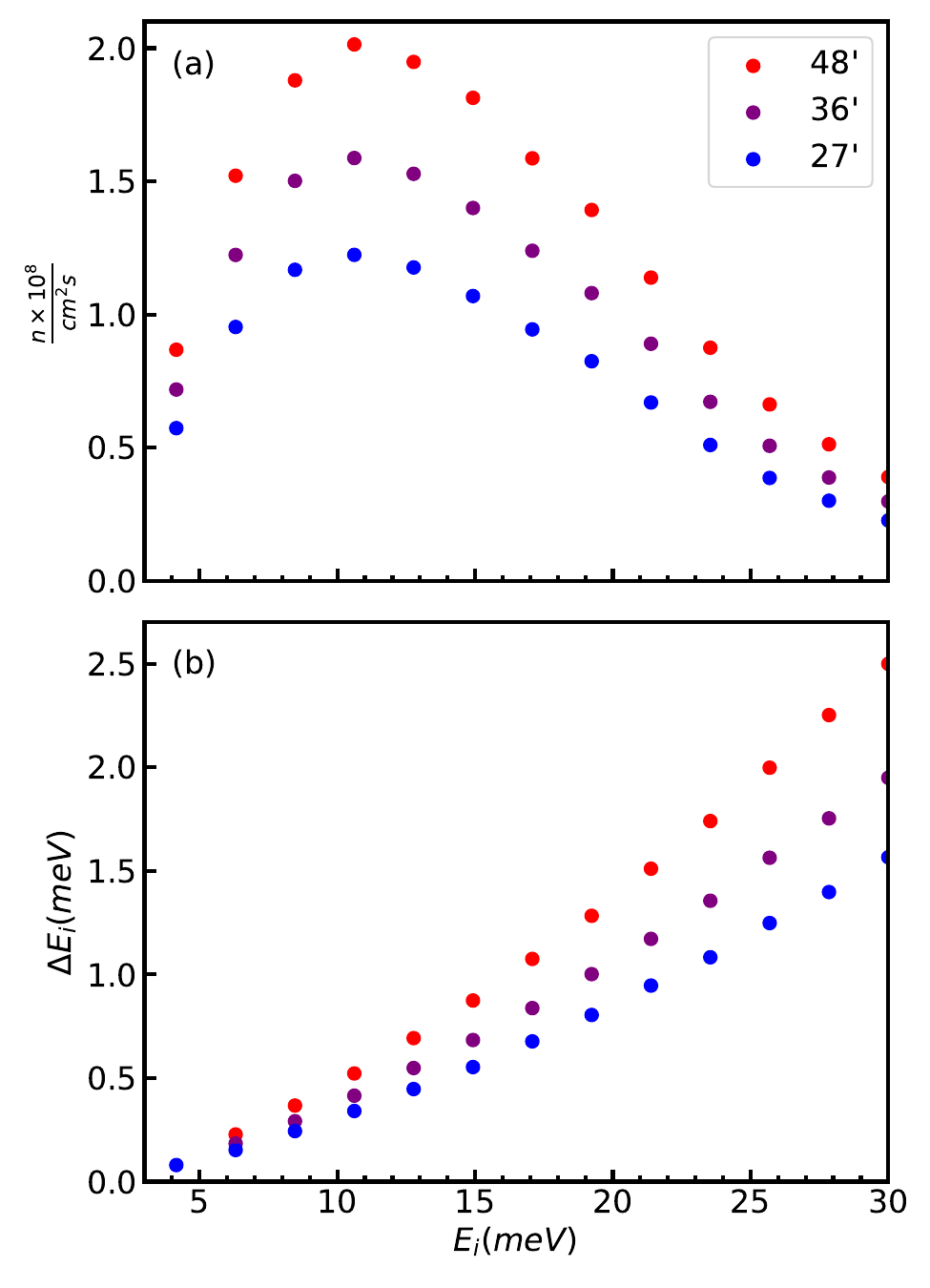}
    \caption{The comparison of three different mosaics. (a) The flux and (b) the FWHM of the energy distribution on a 2 cm x 2 cm sample as a function of incident energy for each mosaic value.  Though the intensity on the sample increases for larger mosaics, the incident energy distribution spreads wider as well. This second effect would adversely affect the overall instrumental resolution. }
    \label{fig:mos_comp}
\end{figure}

A 2 cm by 2 cm region was integrated over this monitor for each $E_i$ value to provide an intensity comparison. To compare the width in each direction, a fit to a Gaussian was performed for each of the two cuts. An example is provided in the upper left and bottom right panels of Fig.~\ref{fig:spot_xmp}. The results for the intensity are plotted in panel (a) of Fig.~ \ref{fig:mos_comp}. 
An increase in beam size is seen from these fits as well, which could be controlled by slits before the sample. However this wider beam would be a concern for background from scatter off the sample environment and air. 
The $E_i$ distribution on the sample is of more concern. This was checked by placing a spectrum monitor at the sample location of the simulated instrument.  For each incident energy a Gaussian was fit to the distribution.  The FWHM of this distribution as a function of $E_i$ is plotted for each mosaic in panel (b) of Fig.~ \ref{fig:mos_comp}. Note that the energy distribution broadens significantly with mosaics above 27'. Such broadening would adversely affect the resolution of the instrument.  Therefore, the 27' crystal mosaic is the best choice for the monochromator crystals.

\begin{figure}
    \centering
    \includegraphics[width=0.5\textwidth]{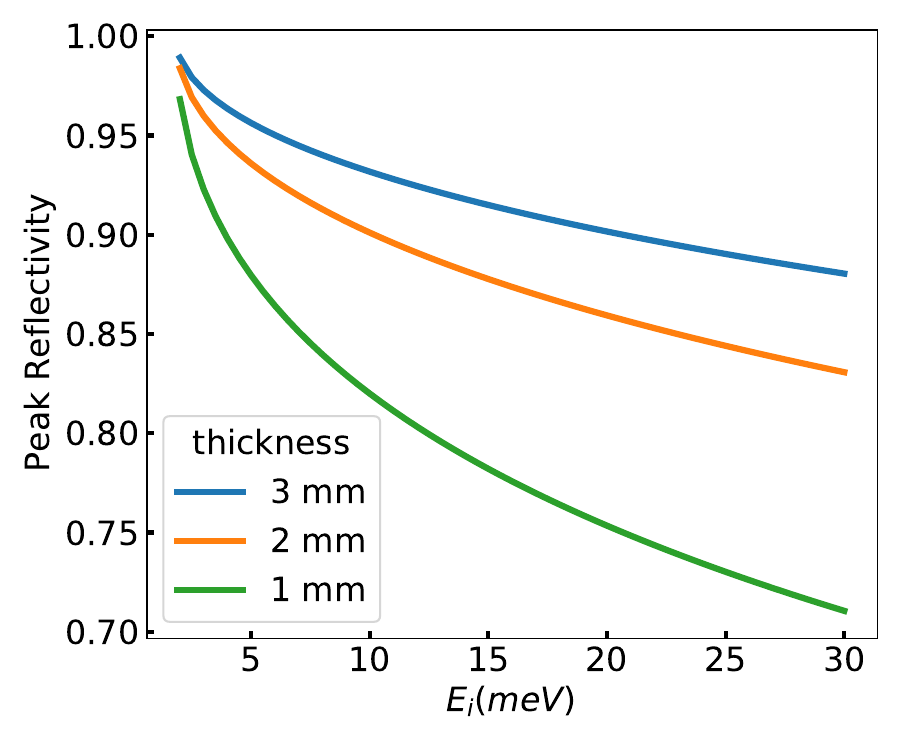}
    \caption{Reflectivity as a function of incident energy for different thickness of PG crystals.}
    \label{fig:mon_thick}
\end{figure}

Finally to complete the optimization of the graphite crystals, standard reflectivity calculations \cite{shirane2002neutron,RISTE1969197,freund:1979,sears1989neutron} were performed to determine the appropriate thickness. 
The results are shown in Fig.~\ref{fig:mon_thick}. Notice that for the 2 mm thick crystals, $\sim$90\% reflectivity is preserved through the operational range of the instrument.
Choosing 1 mm thick crystals impacts the reflectivity significantly. While 3 mm thick crystals provide even better reflectivity, the improved performance over their 2 mm thick counterparts is minimal. 
This crystal thickness is typical of other cold triple axis instruments\cite{Rodriguez_2008,SKOULATOS2011100,doi:10.1080/10448632.2015.1057050,SCHMALZL201689}  More generally for monochromators on cold instruments, thinner crystals may be optimal\cite{Freund:in5061}. But instruments that bridge the cold to thermal range, like the one designed here, require the higher reflectivity from the thicker crystals. 

\section{Neutron Velocity Selector Performance}
\begin{figure}
\includegraphics[width=0.45\textwidth]{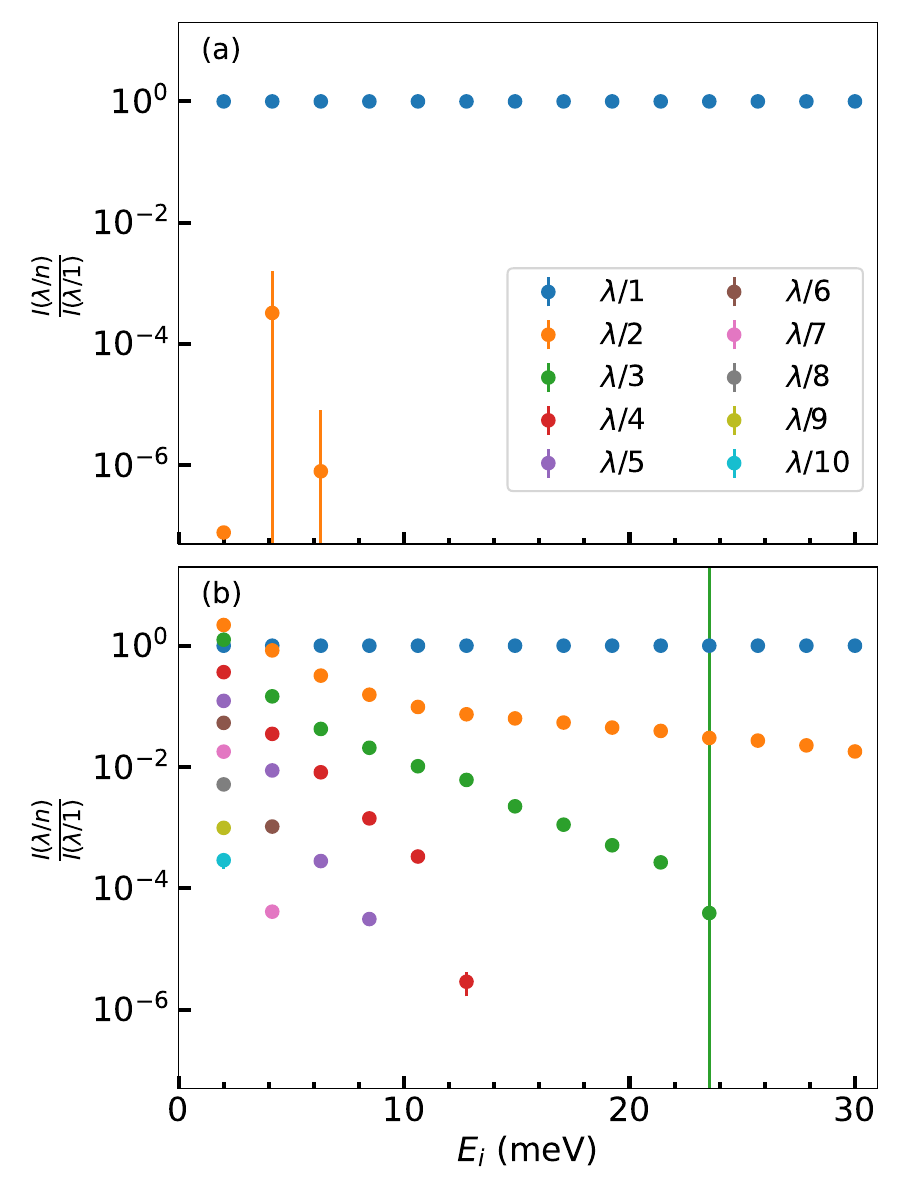}
\caption{The effectiveness of the NVS. (a) With the NVS, only $\lambda /2 $ is observed at a level of at least 4 orders of magnitude below the primary signal. (b)Without the NVS, many $\lambda /n $ reflections are present in significant amounts.}
\label{fig:order_reject}
\end{figure}

A neutron velocity selector is a device with a set of channels that run along the beam direction.  These channels rotate around an axis near parallel to the beam and have a twist angle that follows the path of a defined bandwidth of neutrons in the rotating frame of this rotor \cite{doi:10.1063/1.1770658,Brugger1965}. 
Such a device limits the transmitted bandwidth of the neutron beam, while still transmitting a beam that is continuous in time. 
The speed of the rotor and, the angle of its rotation axis with respect to the beam, can be adjusted to only permit the primary wavelength of neutrons through the device.
The planned device is a standard from Airbus with a twist angle of $\alpha = 22.05^\circ$ and a length of 250 mm\cite{bertzold1989}.
It is desirable to place the neutron velocity selector (NVS) as far upstream from the monochromator as possible. At such a location, the background from neutrons scattered from the device can be effectively shielded. 
It will be placed on a translation table so it can be moved out of the beam for maintenance or potentially swapped out for alternative filtering components that would be better matched to a specific experiment. Simulations with and without the NVS were performed to test its effectiveness at higher-order wavelength rejection. Specifically, the simulations with the optimized guide configuration and monochromator were used and an intensity-as-a-function-of-energy monitor was placed at the sample position. For each $E_i$ simulated, $1\times 10^9$ neutrons were propagated. A Gaussian was used to fit each peak in the energy monitor. Then the ratio of the observed peak of higher order to the desired energy was plotted. The results are shown in Fig.~\ref{fig:order_reject}.

Without the neutron velocity selector in place, for the lowest $E_i$ values, up to 10 orders of multiple reflection are observed. Furthermore, in some cases the higher order reflections may be more intense than the 1st order. This shows the necessity of a filter.  With the neutron velocity selector in place, higher orders are rejected to a level of a least $10^{-4}$.  However, note that even with this large-event simulation, the statistics on the higher orders that make it through the NVS are so small that the uncertainty is large.

\section{Flux on sample}
\begin{figure}
\includegraphics[width=0.5\textwidth]{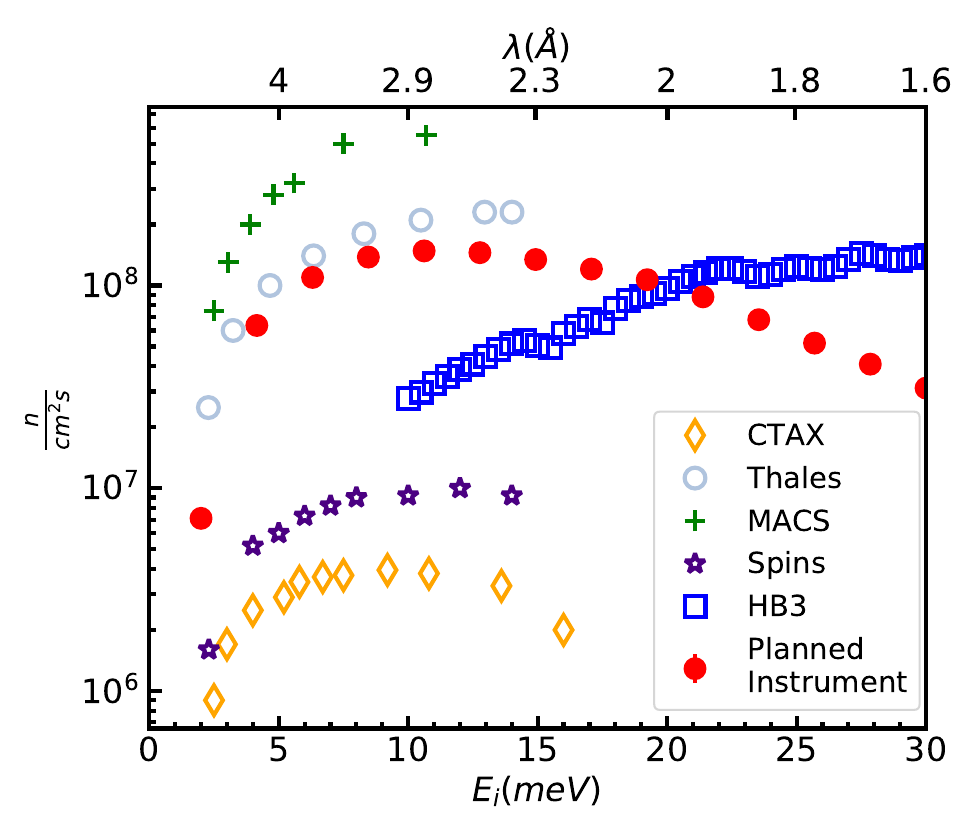}
\caption{The neutron flux on the sample position for a new high-flux cold triple axis spectrometer planned for HFIR as compared to various existing triple axis spectrometers.}
\label{fig:FinalFlux}
\end{figure}

To understand the neutron beam on the sample, an image monitor was placed at the sample position and simulations were preformed for specific incident energies in a fully focused mode. 
Just like for the mosaic study, for each $E_i$ the beam spot monitor was loaded and 2 cm cuts in both the vertical and horizontal directions were extracted. Each of these cuts were fit to a Gaussian to find the number of neutrons in a 2 cm $\times$ 2 cm spot size. 
The fitting results as a function of $E_i$ are normalized by the spot size to provide the plotted flux values (the solid red points) in Fig.~\ref{fig:FinalFlux}. 
For comparison, flux results from other triple axis spectrometers are also plotted, including HB-3 \cite{SongxueDiscussion} and CTAX \cite{TaoDiscussion} at HFIR, MACS and SPINS \cite{Rodriguez_2008} at the National Institute of Standards and Technology, and ThALES \cite{doi:10.1080/10448632.2015.1057050} at the Institute Laue-Langevin. Clearly, the modern cold triple axis spectrometer at HFIR will have a comparable flux to other world-class triple axis spectrometers around the world.

\section{Conclusions}
To summarize, this work presents a detailed design study aimed at optimizing the guide system and monochromator for a modern cold triple axis spectrometer at the High Flux Isotope Reactor. The performance with a neutron velocity selector in place is excellent and the design is well-suited for both traditional single-analyzer and multiplexed secondary spectrometers. 
Incident beamline simulations, performed with the optimized model described here, were used to perform a detailed study of such a multiplexed backend following the CAMEA design \cite{desai2023montecarlo,10.1063/1.4943208,10.1063/5.0128226}.
There is an excellent opportunity to build a world-class reactor-based spectroscopy instrument at ORNL that will serve the condensed matter physics community well for decades to come. 

\section{Acknowledgements}
We are grateful for useful discussions with M. Hoffman, L. Robertson and L. Crow on building constraints. 
We had insightful discussions on instrument design concepts with T. Hong, C. Schanzer, L. Crow, and G. Sala.  
We thank T. Huegle for help with the tally components for determining the guide $m$ value. 
We thank M. Frost for early access to the source parameters. 
We are grateful to S. Chi for providing flux measurements for the HB-3 spectrometer.  
B. McHargue provided current details on the neutron velocity selector design.
The work of M. Daum was supported by the U.S. Department of Energy (USDOE), Office of Science, Office of Workplace Development for Teachers and Scientists, Office of Science Graduate Student Research (SCGSR) program.
The SCGSR is administered by the Oak Ridge Institute for Science and Education for the DOE under contract number DE-SC0014664.
This research used resources of the High Flux Isotope Reactor, which is an  USDOE Office of Science User Facility.


 \bibliographystyle{elsarticle-num} 
 \bibliography{manta}


\end{document}